\documentclass[reprint,NumberedRefs]{JASA}

\usepackage[shortlabels]{enumitem}

\usepackage{algpseudocode}

\usepackage{tikz}
\usetikzlibrary{calc}
\usetikzlibrary{arrows}
\usetikzlibrary{spy}
\usepackage{relsize}
\usepackage{wrapfig}
\usepackage{eurosym}
\usepackage[siunitx]{circuitikz}
\usepackage{amssymb}
\usepackage{esint}
\usepackage{mathrsfs}
\usepackage{amsthm}
\usepackage{amscd}
\usepackage{listings}
\usepackage{booktabs}
\usepackage{multirow}
\usepackage{textcomp}

\newcommand{\vect}[1]{\boldsymbol{#1}}

\newcommand{\matr}[1]{\mathbf{#1}}

\newcommand{\vrup}[0]{\mathbf{x}}

 \newcommand{\unitup}[1]{\vect{\hat{#1}}}

\newcommand{\junk}[1] {}

\def\XXint#1#2#3{{\setbox0=\hbox{$#1{#2#3}{\int}$}
\vcenter{\hbox{$#2#3$}}\kern-.5\wd0}}

\newcommand*\widebar[1]{%
  \hbox{%
    \vbox{%
      \hrule height 0.5pt 
      \kern0.3ex
      \hbox{%
        \kern-0.05em
        \ensuremath{#1}%
        \kern-0.05em
      }%
    }%
  }%
}

\usepackage[siunitx]{circuitikz}
\usepackage{etoolbox}
\makeatletter

\pgfcircdeclarebipole{}
    {\ctikzvalof{bipoles/tline/height}}{tline}
    {\ctikzvalof{bipoles/tline/height}}
    {\ctikzvalof{bipoles/tline/width}}
{   
    \pgfpointdiff{\pgfpointanchor{\ctikzvalof{bipole/name}start}{center}}
                 {\pgfpointanchor{\ctikzvalof{bipole/name}end}{center}}
    \pgfmathparse{veclen(\the\pgf@x,\the\pgf@y)}
    \pgftransformxshift{\pgfmathresult/2-30pt}
    \pgf@circ@res@left=\dimexpr-\pgfmathresult pt+40pt\relax
    \pgf@circ@res@step=.2\pgf@circ@res@right 
    \pgfsetlinewidth{\pgfkeysvalueof{/tikz/circuitikz/bipoles/thickness}\pgfstartlinewidth}
    \pgfpathellipse{\pgfpoint{\pgf@circ@res@right-\pgf@circ@res@step}{0}}
                   {\pgfpoint{\pgf@circ@res@step}{0}}
                   {\pgfpoint{0}{-\pgf@circ@res@up}}
    \pgfpathmoveto{\pgfpoint{\pgf@circ@res@right-\pgf@circ@res@step}{\pgf@circ@res@up}}
    \pgfpathlineto{\pgfpoint{\pgf@circ@res@left+\pgf@circ@res@step}{\pgf@circ@res@up}}
    \pgfpatharc{-90}{90}{-\pgf@circ@res@step and -\pgf@circ@res@up}
    \pgfpathlineto{\pgfpoint{\pgf@circ@res@right-\pgf@circ@res@step}{\pgf@circ@res@down}}
    \pgfsetfillcolor{white}
    \pgfusepath{draw,fill}
    \pgfpathmoveto{\pgfpoint{\pgf@circ@res@right-2*\pgf@circ@res@step}{0pt}}
    \pgfpathlineto{\pgfpoint{\pgf@circ@res@right}{0pt}}
    \pgfusepath{draw}
}

\allowdisplaybreaks

\usepackage{pgfplots}
\usepackage{soul}

\begin{document}

\title[Wigner-Smith Time Delay Matrix for Acoustic Scattering: Theory and Phenomenology]{Wigner-Smith Time Delay Matrix for Acoustic Scattering: Theory and Phenomenology}
\author{Utkarsh R. Patel}
\author{Yiqian Mao}
\author{Eric Michielssen}
\affiliation{Department of Electrical Engineering and Computer Science,  University of Michigan, Ann Arbor, Michigan 48109-2122, United States}

\date{\today} 

\begin{abstract}
The Wigner-Smith (WS) time delay matrix relates a lossless system's scattering matrix to its frequency derivative. First proposed in the realm of quantum mechanics to characterize time delays experienced by particles during a collision, this article extends the use of WS time delay techniques to acoustic scattering problems governed by the Helmholtz equation. Expression for the entries of the WS time delay matrix involving renormalized volume integrals of energy densities are derived, and shown to hold true independent of the scatterer's geometry, boundary condition (sound-soft or sound-hard), and excitation. Numerical examples show that the eigenmodes of the WS time delay matrix describe distinct scattering phenomena characterized by well-defined time delays.
\end{abstract}


\maketitle

\section{\label{sec:1} Introduction}

The concept of the Wigner-Smith (WS) time delay matrix $\matr{Q}$ was proposed by Felix Smith to characterize time delays experienced by quantum particles during a collision~\cite{Smith_1960}. 
Specifically, Smith showed that the matrix
\begin{align}
    \matr{Q} = j \matr{S}^\dag \frac{\partial \matr{S}}{\partial k}
    \label{eq:WS_2}
\end{align}
where $\matr{S}$ is the scattering matrix of a lossless potential well, $\partial/\partial k$ denotes the derivative w.r.t. wavenumber $k$, and $\,^\dag$ represents the adjoint operation,
fully characterizes the time particles dwell in a system. 

Since its introduction in 1960, the WS time delay matrix has found many applications in quantum mechanics, photonics, and electromagnetics.
In quantum mechanics, the WS time delay matrix $\matr{Q}$ has been used to study photoionization and photoemission time delays~\cite{Gallmann_2017, Hockett_2016}, to analyze traversal times during quantum tunneling~\cite{Buttiker_1982, Wardlaw_1988}, and to characterize quantum mechanical decay mechanisms~\cite{Dittes_2000}. 
In photonics, $\matr{Q}$ has been used to study wave propagation in multimode fibers~\cite{Carpenter_2015}, to shape light flow in disordered media and complex cavities~\cite{Brandstotter_2019, Gerardin_2016, Bohm_2018}, and to optimize light storage in highly scattering media~\cite{Durand_2019}. 
In electromagnetics, WS time delays have been used for focusing energy in a microwave cavity~\cite{Amb2017focus}, to characterize 
group delays of fields interacting with a two-port waveguide~\cite{Winful_2003}, to decompose electromagnetic fields in terms of modes with well-defined time delays~\cite{TAP1_WS}, and to characterize the frequency sensitivity of antenna input impedances~\cite{TAP2_WS}.

Applications of WS time delay concepts to acoustic scattering problems have been few and far in between, however. Exceptions include the use of $\matr{Q}$ to study acoustic resonances in fluid-loaded elastic plates~\cite{Frank2006}, the study of scattering from two-dimensional fluid slabs in an elastic medium~\cite{Rembert2007four}, and the characterization of elastic wave propagation in disordered media~\cite{gerardin2014full}.

This work, along with its companion paper ``Wigner-Smith Time Delay Matrix for Acoustic Scattering: Computational Aspects'', extends the authors' work~\cite{TAP1_WS} on WS methods for Maxwell's equations (electromagnetics) to the Helmholtz equation (acoustics). 
The WS time delay matrix of a lossless acoustic scatterer possesses several interesting properties that carry over wholesale from its electromagnetic counterpart, and are summarized next -- the reader is referred to the section on ``WS Theory: Systems Perspective'' in reference~\cite{TAP1_WS} for a detailed derivation and justification of the statements that follow. 
\begin{itemize}[leftmargin=*]
\item First, the WS time delay matrix $\matr{Q}$ is Hermitian and hence its diagonal elements are purely real. This property can be derived by using the symmetry and unitary properties of the scattering matrix, i.e. $\matr{S}^T = \matr{S}$ and $\matr{S}^\dag \matr{S} = \matr{I}$, where $\matr{I}$ is an identity matrix and $\,^T$ denotes the transpose operation. 
\item Second, the $n$-th diagonal element of $\matr{Q}$ represents the average time delay experienced by an incoming narrowband wave packet that enters the system via its $n$-th port prior to exiting via any other port. 
\item Third, $\matr{Q}$ and $\matr{S}$ can be simultaneously diagonalized as $\matr{Q} = \matr{W} \widebar{\matr{Q}} \matr{W}^\dag$  and $\matr{S} = \matr{W}^* \widebar{\matr{S}} \matr{W}^\dag$.  $\matr{W}$ is a unitary matrix whose columns are eigenvectors of $\matr{Q}$ and are often called ``WS modes''. $\widebar{\matr{Q}}$ is a real diagonal matrix that collects $\matr{Q}$'s eigenvalues, often called ``WS time delays'' or ``WS dwell times''. 
\item Fourth, the simultaneous diagonalization of $\matr{Q}$ and $\matr{S}$ implies that WS modes are decoupled and are characterized by well-defined time delays. In other words, an incoming field composed of narrowband wave packets that enter the system weighed by the entries of $\matr{Q}$’s $n$-th eigenvector exit the system with a time delay characterized by $\matr{Q}$’s $n$-th eigenvalue. 
\end{itemize}

This paper elucidates the definition and use of the WS time delay matrix for acoustic scattering problems governed by the Helmholtz equation. Its  contributions are twofold.  
\begin{itemize}[leftmargin=*]
\item First, it demonstrates that the elements of $\matr{Q}$ can be expressed as renormalized volume integrals of energy density-like quantities involving fluid potentials. The derived volume integral expressions for elements of $\matr{Q}$ provide an alternative to Eqn.~\eqref{eq:WS_2}, allowing $\matr{Q}$'s computation without knowledge of $\matr{S}$ and $\partial \matr{S}/\partial k$. When used in conjunction with Eqn.~\eqref{eq:WS_2}, the volume integral expression for $\matr{Q}$ can be used to compute $\partial \matr{S}/\partial k$ provided $\matr{S}$ is known. 

\item Second, it demonstrates via numerical experimentation that WS modes can be used to untangle various wave components, including those associated with corner/edge diffraction, surface waves, ballistic scattering, and resonant modes, and that each of these is characterized by distinct and well-defined time delays/dwell times and stored energies.
\end{itemize}
The two topics above are detailed in Secs.~II and Sec.~III below.
Conclusions and avenues for future research are provided in Sec.~IV. Throughout this paper, the superscript $'$ denotes ${\partial}/{\partial} k$.

\section{WS Theory for Acoustic Scattering}
\label{sec:3}

\begin{figure}
\centering
\includegraphics{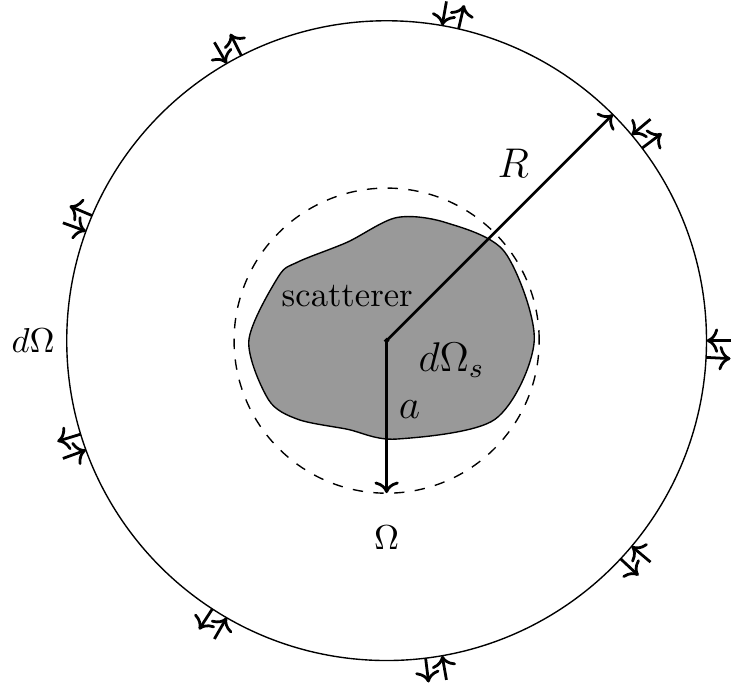}
\caption{Scattering system excited through free-space port defined on sphere of radius $R$.}
\label{fig:sample_scatterer}
\end{figure}

This section shows that the WS time delay matrix for acoustic scattering problems governed by the Helmholtz equation can be computed directly by integrating energy-like densities involving the total fluid velocity potential over $\mathbb{R}^3$.

\subsection{Setup}
Consider a closed scatterer $\Omega_s$ with surface $d\Omega_s$ that resides in a lossless and homogeneous medium with (real) wavenumber $k$ and is circumscribed by a sphere of radius $a$ centered at the origin (Fig.~\ref{fig:sample_scatterer}).
Let $\Omega$ and $d\Omega$ denote the volume and surface of a sphere of radius $R \gg a$ and $kR \gg 1$, also centered about the origin. 
The fluid velocity potential $\phi(\vrup)$ in a source-free medium $\mathbb{R}^3 \backslash \Omega_s$ satisfies the Helmholtz equation
\begin{align}
    \nabla^2 \phi(\vrup) + k^2 \phi(\vrup) = 0
    \label{eq:Helmholtz}
\end{align}
subject to the Dirichlet boundary condition
\begin{align}
    \phi(\vrup) = 0, \quad \vrup \in d\Omega_s
    \label{eq:soundsoft_bc}
\end{align}
for a sound-soft scatterer or the Neumann boundary condition
\begin{align}
    \frac{\partial \phi(\vrup)}{\partial n(\vrup)} = 0, \quad \vrup \in d\Omega_s
    \label{eq:soundhard_bc}
\end{align}
for a sound-hard scatterer; here $\partial/\partial n(\vrup)$ denotes the outward normal derivative on $d\Omega_s$. 

On $d\Omega$, the velocity potential may be expanded in terms of $M$ orthonormal modes ${\cal X}_p(\theta,\phi)$.
While there exist many different choices for modes ${\cal X}_p(\theta,\phi)$, this paper adopts a specific realization based on spherical harmonics ${\cal X}_p(\theta,\phi)$ with $p=(l,m)$, where $l=0,\hdots, l_{max}$, $m = -l,\hdots, l$, detailed in Appendix~\ref{Appdix:modes}; $M = (l_{max}+1)^2$, $l_{max} = k a + c(ka)^{1/3}$ and typically $2 < c< 4$ \cite{Wiscombe1980}; these modes are orthonormal, i.e.
\begin{align}
    \int_{0}^{2\pi} \int_{0}^\pi {\cal X}_{p}(\theta,\phi) {\cal X}_{q}^*(\theta,\phi) \sin \theta d\theta d\phi = \delta_{p,q} \label{eq:orthonormal}
\end{align}
where $\delta_{l,l'}$ is a Kronecker delta function.

Assume that the scatterer is excited by a radially-travelling \textit{incoming} ($i$) unit-power field $\phi_p^i(\vrup)$. Using large argument approximations, this field on $d\Omega$ reads
\begin{align}
    \phi_{p,\infty}^{i}(\vrup) = {\cal X}_p(\theta,\phi) \frac{e^{jk r}}{r}\,. \label{eq:phip_i}
\end{align}
Next, let $\phi_p(\vrup)$ denote the \textit{total} field throughout $\Omega$ due to the excitation in Eqn.~\eqref{eq:phip_i}. Near $d\Omega$, this field is
\begin{align}
    \phi_{p,\infty}(\vrup) = \phi_{p,\infty}^i(\vrup) + \phi_{p,\infty}^o(\vrup)
    \label{eq:phi_total}
\end{align}
where $\phi_{p,\infty}^o(\vrup)$, the \textit{outgoing} field near $d\Omega$, can be expressed as
\begin{align}
    \phi_{p,\infty}^o(\vrup) &= \sum_{m=1}^{M} \matr{S}_{mp} {\cal X}_{m}^*(\theta,\phi) \frac{e^{-jkr}}{r}\,.
    \label{eq:phi_po}
\end{align}
That is, $\phi_{p,\infty}^o(\vrup)$ is the superposition of $M$ outgoing modes weighed by wavenumber-dependent scattering coefficients $\matr{S}_{mp}$.
The above construct guarantees that $\matr{S}$ is symmetric and unitary, i.e.
\begin{subequations}
\begin{align}
    \matr{S}^\dag \matr{S} &= \matr{I}\\
    \matr{S}^T &= \matr{S} \,.
\end{align}
\end{subequations}

In preparation of the derivation of the WS relationship, consider two velocity potentials, $\phi_p(\vrup)$ and $\phi_q(\vrup)$.
Taking the derivative w.r.t $k$ of Eqn.~\eqref{eq:Helmholtz} with $\phi(\vrup) = \phi_p(\vrup)$ yields
\begin{align}
    \nabla^2 \phi_p'(\vrup) + 2k  \phi_p(\vrup) + k^2 \phi_p'(\vrup) &= 0\,. \label{eq:1c}
\end{align}
Multiplying the conjugate of Eqn.~\eqref{eq:Helmholtz} with $\phi(\vrup) = \phi_q(\vrup)$ by $\phi_p'(\vrup)$ and subtracting from it Eqn.~\eqref{eq:1c} multiplied by $\phi_q^*(\vrup)$ yields
\begin{align}
    \frac{1}{2 k} (\phi_p'(\vrup) \nabla^2 \phi_q^*(\vrup) &- \phi_q^*(\vrup) \nabla^2 \phi_p'(\vrup)) \nonumber \\
    &= \phi_p(\vrup) \phi_q^*(\vrup)\,. \label{eq:2a}
\end{align}
Using the identity $\psi \nabla^2 \gamma - \gamma \nabla^2 \psi = \nabla \cdot \left(\psi \nabla \gamma - \gamma \nabla \psi \right)$, Eqn.~\eqref{eq:2a} simplifies to
\begin{align}
    \frac{1}{2k} \nabla \cdot \left( \phi_p'(\vrup) \nabla \phi_q^*(\vrup) - \phi_q^*(\vrup) \nabla \phi_p'(\vrup) \right) \nonumber \\
    =  \phi_p(\vrup) \phi_q^*(\vrup) \,. \label{eq:3a}
\end{align}
Next, the RHS of Eqn.~\eqref{eq:3a} is rewritten using Eqn.~\eqref{eq:Helmholtz} with $\phi(\vrup) = \phi_q^*(\vrup)$ and the vector identity $\phi \nabla^2 \gamma = \nabla \cdot (\phi \nabla \gamma) - \nabla \gamma \cdot \nabla \phi$ to obtain 
\begin{align}
 \frac{1}{2k} \nabla \cdot &\left( \phi_p'(\vrup) \nabla \phi_q^*(\vrup) - \phi_q^*(\vrup) \nabla \phi_p'(\vrup) \right) \nonumber \\
    =& \frac{1}{2} \phi_p(\vrup) \phi_q^*(\vrup) \nonumber \\
    &- \frac{1}{2 k^2} \nabla \cdot (\phi_p(\vrup)  \nabla \phi_q^*(\vrup) ) \nonumber \\
    &+ \frac{1}{2 k^2} \nabla \phi_q^*(\vrup)  \cdot \nabla \phi_p(\vrup) \,.
    \label{eq:4}
\end{align}
Finally, integrating Eqn.~\eqref{eq:4} over $\Omega$ and using the divergence theorem yields
\begin{align}
\frac{1}{2k} &\int_{d\Omega}  \unitup{r} \cdot \bigg[ \phi_{p,\infty}'(\vrup) \nabla \phi_{q,\infty}^*(\vrup)  - \phi_{q,\infty}^*(\vrup) \nabla \phi_{p,\infty}'(\vrup)  \nonumber \\
&\quad +  \frac{1}{k} \nabla \phi_{q,\infty}^*(\vrup) \phi_{p,\infty}(\vrup)\bigg]   d\vrup \nonumber \\
    &=  \frac{1}{2} \int_{\Omega} \phi_p(\vrup) \phi_q^*(\vrup) d\vrup \nonumber \\
    &\quad + \frac{1}{2k^2} \int_{\Omega} \nabla \phi_q^*(\vrup) \cdot \nabla \phi_p(\vrup) d\vrup\,. 
    \label{eq:5}
    \end{align}
The above equation assumes a hard or soft-source boundary condition to eliminate the integration over $d\Omega_s$ on the LHS.

\subsection{WS Relationship}
The evaluation of the LHS of Eqn.~\eqref{eq:5} requires expressions for $\phi_{p,\infty}'(\vrup)$, $\unitup{r} \cdot \nabla \phi_{p,\infty}'(\vrup)$, $\phi_{q,\infty}^*(\vrup)$, and $\unitup{r} \cdot \nabla \phi_{q,\infty}^*(\vrup)$.
Substituting Eqns.~\eqref{eq:phip_i} and~\eqref{eq:phi_po} into Eqn.~\eqref{eq:phi_total} and differentiating w.r.t. $k$ yields
\begin{align}
    \phi_{p,\infty}'(\vrup) &= j {\cal X}_p(\theta,\phi) e^{jkr} \nonumber \\
    &\quad + \sum_{n=1}^{M} \matr{S}_{np}' {\cal X}_n^*(\theta,\phi) \frac{e^{-jkr}}{r} \nonumber \\
    &\quad - j\sum_{n=1}^{M} \matr{S}_{np} {\cal X}_n^*(\theta,\phi) e^{-jkr} \,.
    \label{eq:phip_dk}
\end{align}
Taking the gradient of Eqn.~\eqref{eq:phip_dk}, dotting the result with $\unitup{r}$, and using the large argument behavior of $r\rightarrow \infty$ produces
\begin{align}
    \unitup{r} &\cdot \nabla \phi_{p,\infty}'(\vrup) \nonumber \\
    &= -k {\cal X}_p(\theta,\phi) e^{jkr} \nonumber \\
    &\quad + \sum_{n=1}^{M} \matr{S}_{np}' {\cal X}_n^*(\theta,\phi) \frac{e^{-jkr}}{r}\left(-jk\right) \nonumber \\
    &\quad - k \sum_{n=1}^{M} \matr{S}_{np} {\cal X}_n^*(\theta,\phi) e^{-jkr}.
    \label{eq:gradphip_dk}
\end{align}
Likewise, substituting Eqns.~\eqref{eq:phip_i} and~\eqref{eq:phi_po} into Eqn.~\eqref{eq:phi_total} while changing $p$ to $q$, and conjugating the result yields
\begin{align}
    \phi_{q,\infty}^*(\vrup) &= {\cal X}_{q}^*(\theta,\phi) \frac{e^{-jkr}}{r} \nonumber \\
    &\quad +  \sum_{m=1}^{M} \matr{S}_{mq}^* {\cal X}_{m}(\theta,\phi) \frac{e^{jkr}}{r}\,.
    \label{eq:phiq_conj}
\end{align}
Finally, taking the gradient of Eqn.~\eqref{eq:phiq_conj}, dotting it with $\unitup{r}$ and using the large argument approximation for $r$ results in
\begin{align}
    \unitup{r} \cdot \nabla \phi_{q,\infty}^*(\vrup) &= {\cal X}_q^*(\theta,\phi) \frac{e^{-jkr}}{r} \left(-jk\right) \nonumber \\
    &+ \sum_{m=1}^{M} \matr{S}_{mq}^* {\cal X}_m(\theta,\phi) \frac{e^{jkr}}{r} \left(jk\right)\,.
    \label{eq:gradphiq_conj}
\end{align}
Next, substituting Eqns.~\eqref{eq:phip_dk}, \eqref{eq:gradphip_dk}, \eqref{eq:phiq_conj}, and \eqref{eq:gradphiq_conj} into the LHS of Eqn.~\eqref{eq:5} and evaluating the resulting integral as detailed in Appendix~\ref{Appdix:sur_int} yields
\begin{align}
    \frac{1}{2k} &\int_{d\Omega}  \unitup{r} \cdot \bigg[ \phi_{p,\infty}'(\vrup) \nabla \phi_{q,\infty}^*(\vrup)  - \phi_{q,\infty}^*(\vrup) \nabla \phi_{p,\infty}'(\vrup)  \nonumber \\
&\quad +  \frac{1}{k} \nabla \phi_{q,\infty}^*(\vrup) \phi_{p,\infty}(\vrup)\bigg]   d\vrup \nonumber \\
&= 2R \delta_{p,q}   + j  \sum_{m=1}^{M} \matr{S}_{mq}^* \matr{S}_{mp}'.
\label{eq:5LHS}
\end{align}
By using Eqn.~\eqref{eq:5LHS}, Eqn.~\eqref{eq:5} simplifies to
\begin{align}
    \widetilde{\matr{Q}}_{qp} = 2 R \delta_{p,q} + j \sum_{m=1}^{M} \matr{S}_{mq}^* \matr{S}_{mp}'
    \label{eq:WS_derive1}
\end{align}
where $\widetilde{\matr{Q}}_{qp}$ reads
\begin{align}
    \widetilde{\matr{Q}}_{qp} &= \frac{1}{2} \int_{\Omega} \phi_p(\vrup) \phi_q^*(\vrup) d\vrup \nonumber \\
    &\quad + \frac{1}{2k^2} \int_{\Omega} \nabla \phi_q^*(\vrup) \cdot \nabla \phi_p(\vrup) d\vrup.
    \label{eq:5RHS}
\end{align}
To obtain a WS relationship that is independent of $R$, consider the quantity $\widetilde{\matr{Q}}_{qp,\infty}$ obtained by replacing $(\phi_p,\nabla \phi_p)$ and $(\phi_q,\nabla \phi_q)$ by $(\phi_{p,\infty}, \nabla \phi_{p,\infty})$ and $(\phi_{q,\infty}, \nabla \phi_{q,\infty})$, i.e.
\begin{align}
    \widetilde{\matr{Q}}_{qp,\infty} &= \frac{1}{2} \int_{\Omega} \phi_{p,\infty}(\vrup) \phi_{q,\infty}^*(\vrup) d\vrup \nonumber \\
    &\quad + \frac{1}{2k^2} \int_{\Omega} \nabla \phi_{q,\infty}^*(\vrup) \cdot \nabla \phi_{p,\infty}(\vrup) d\vrup.
    \label{eq:5RHS_infty}
\end{align}
The quantities $\phi_{p,\infty}$, $\phi_{q,\infty}$ in Eqn.~\eqref{eq:5RHS_infty} are the same as those in Eqn.~\eqref{eq:phi_total} though their use is extended to all of $\vrup \in \Omega$; a similar interpretation holds true for their gradients.
Evaluating the integral in Eqn.~\eqref{eq:5RHS_infty} yields
\begin{align}
    \widetilde{\matr{Q}}_{qp,\infty} &= 2 R \delta_{p,q} \,.
\end{align}
Finally, subtracting $\widetilde{\matr{Q}}_{qp,\infty}$ from both sides of Eqn.~\eqref{eq:WS_derive1} yields the WS relationship
\begin{align}
    \matr{Q}_{qp} =  j  \sum_{m=1}^{M} \matr{S}_{mq}^* \matr{S}_{mp}' \,,\label{eq:WS1}
\end{align}
where the $(q,p)$-th entry of the WS time delay matrix $\matr{Q}$ is
\begin{align}
    \matr{Q}_{qp} &= \frac{1}{2} \int_{\Omega} \left[\phi_p(\vrup) \phi_q^*(\vrup) - \phi_{p,\infty}(\vrup) \phi_{q,\infty}^*(\vrup) \right]d\vrup \nonumber \\
    &\quad + \frac{1}{2k^2} \int_{\Omega} \big[\nabla \phi_q^*(\vrup) \cdot \nabla \phi_p(\vrup) \nonumber \\
    &\quad \quad \quad - \nabla \phi_{q,\infty}^*(\vrup) \cdot \nabla \phi_{p,\infty}(\vrup) \big] d\vrup\,.
\end{align}
In matrix form, Eqn.~\eqref{eq:WS1} reads
\begin{align}
    \matr{Q}= j\matr{S}^\dag \matr{S}'\,, \label{eq:WS1_matrix}
\end{align}
showing that the entries of the WS time delay matrix are renormalized volume integrals of energy-like densities.

\subsection{Alternative Formulations}

The derivation above casts elements of the WS time delay matrix in terms of integrals involving velocity potentials $\phi_p(\vrup)$ and $\phi_q(\vrup)$, \textit{and} their gradients $\nabla \phi_p(\vrup)$ and $\nabla \phi_q(\vrup)$.
It is often convenient to compute the entries of the WS time delay matrix from just the scalar fields \textit{or} their gradients. 
The first alternative formulation, obtained by integrating Eqn.~\eqref{eq:3a}, reads
\begin{align}
    \matr{Q}^a &= j \matr{S}^\dag \matr{S}'  \label{eq:alternative1}
\end{align}
where 
\begin{align}
    \matr{Q}^a_{qp} = \widetilde{\matr{Q}}_{qp}^a + \frac{j}{2k} (-1)^m \matr{S}_{q,\tilde{p}} - \frac{j}{2k}(-1)^m \matr{S}_{\tilde{p},q}^* 
    \label{eq:Qa}
\end{align}
and
\begin{align}
    \widetilde{\matr{Q}}_{qp}^a &= \int_{\Omega} \left[\phi_{p}(\vrup) \phi_{q}^*(\vrup) - \phi_{p,\infty}(\vrup) \phi_{q,\infty}^*(\vrup)\right] d\vrup\,.
\end{align}  
The derivation of Eqn.~\eqref{eq:Qa} requires use of the conjugation property of spherical harmonics with ${p} = (l,m)$, ${q} = (l',m')$, and $\tilde{p} = (l,-m)$ (see Eqn.~\eqref{eq:conjugation} in Appendix~\ref{Appdix:modes}). 
The second alternative formulation, obtained by integrating two times Eqn.~\eqref{eq:4} and subtracting Eqn.~\eqref{eq:3a}, reads
\begin{align}
    \matr{Q}^b &= j \matr{S}^\dag \matr{S}'  \label{eq:alternative2}
\end{align}
where 
\begin{align}
    \matr{Q}^b_{qp} = \widetilde{\matr{Q}}_{qp}^b - \frac{j}{2k} (-1)^m \matr{S}_{q,\tilde{p}} + \frac{j}{2k}(-1)^m \matr{S}^*_{\tilde{p},q}
\end{align}
and
\begin{align}
    \widetilde{\matr{Q}}_{qp}^b =& \frac{1}{k^2} \int_{\Omega} \bigg[\nabla \phi_{p}(\vrup) \cdot \nabla \phi_{q}^*(\vrup) \nonumber \\
    & \quad \quad - \nabla \phi_{p,\infty}(\vrup) \cdot \nabla \phi_{q,\infty}^*(\vrup)\bigg] d\vrup\,.
\end{align}
It is easily verified that the symmetric formulation in Eqn.~\eqref{eq:WS1_matrix} is obtained by adding Eqns.~\eqref{eq:alternative1} and \eqref{eq:alternative2}.

\section{Illustrative Examples}

This section presents two examples that demonstrate the use of WS methods for decomposing fields interacting with scatterers into WS modes characterized by well-defined time delays. The WS modes for sound-soft and sound-hard scatterers are seen to exhibit different characteristics.
Both examples involve 2D geometries to facilitate visualization of the fields.

\subsection{Thin Rectangular Strip}

\begin{figure*}[htb]
\null \hfill
\figline{\fig{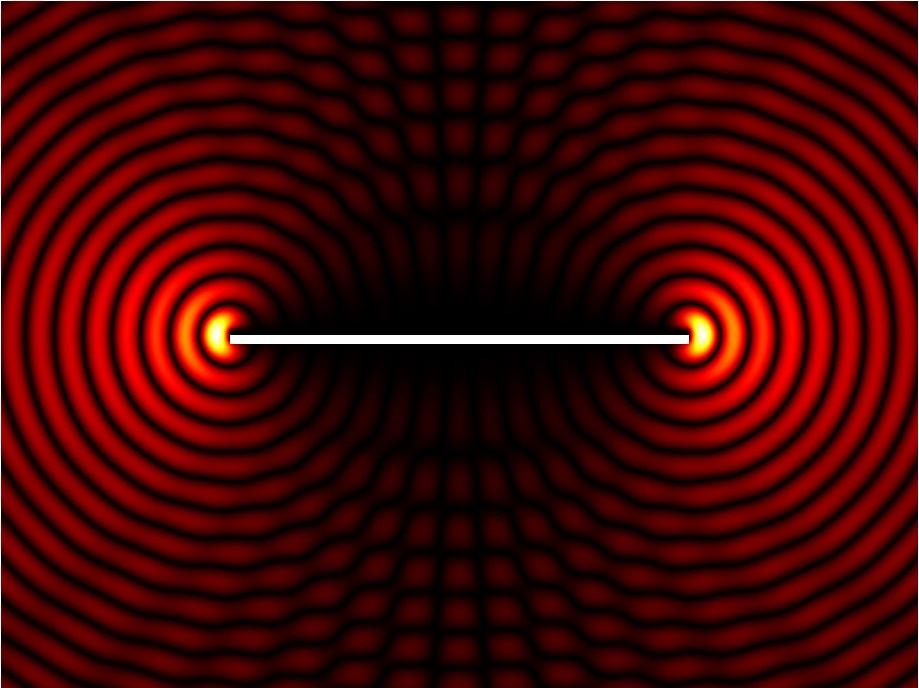}{0.3\textwidth}{(a) WS mode \#1}\label{fig:strip_tm_WS1} 
\fig{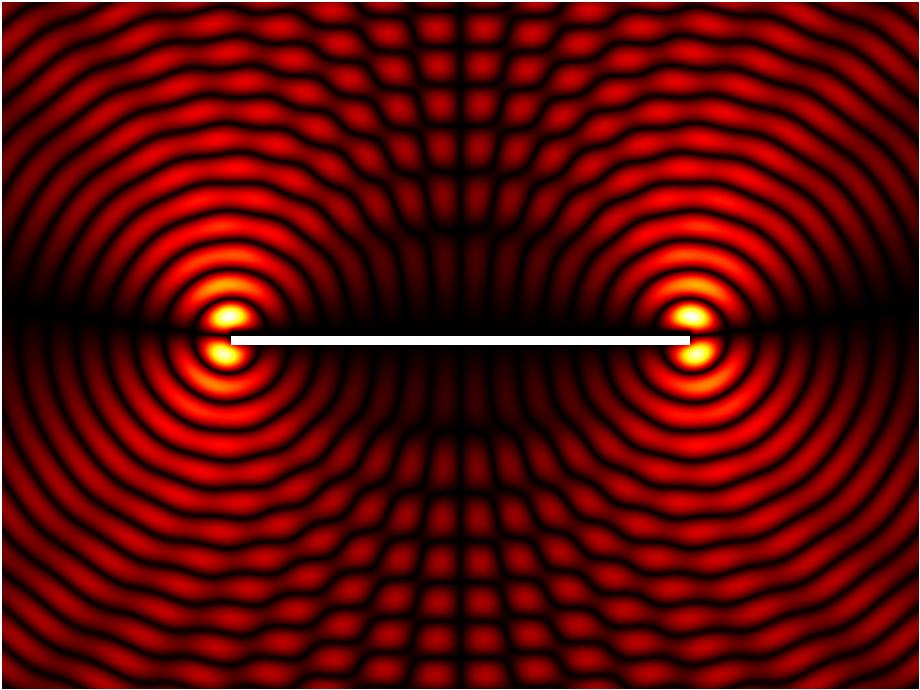}{0.3\textwidth}{(b) WS mode \#3}\label{fig:strip_tm_WS3}
\fig{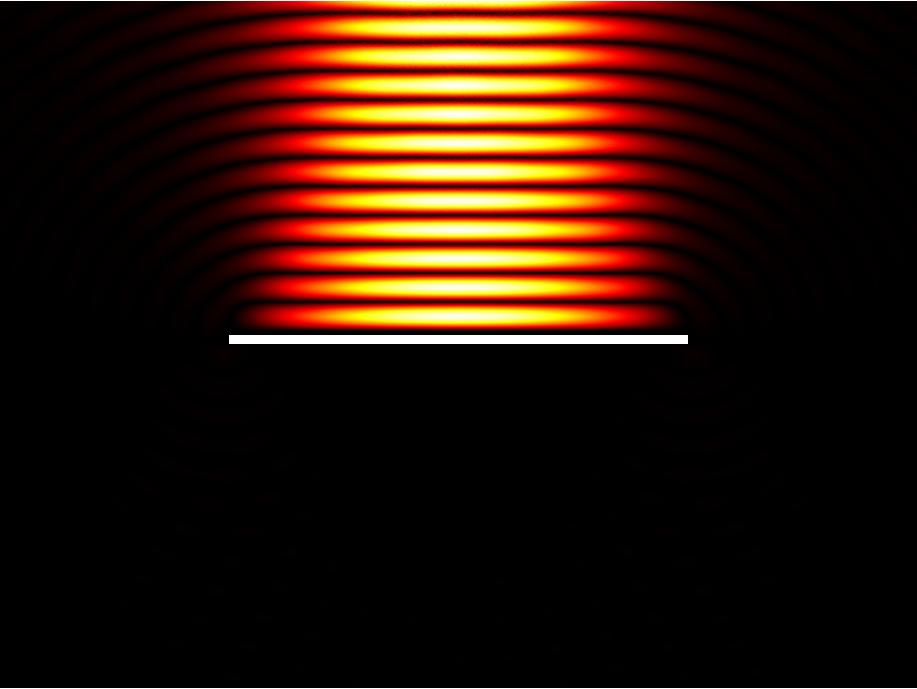}{0.3\textwidth}{(c) WS mode \#5}\label{fig:strip_tm_WS5}}
\figline{\fig{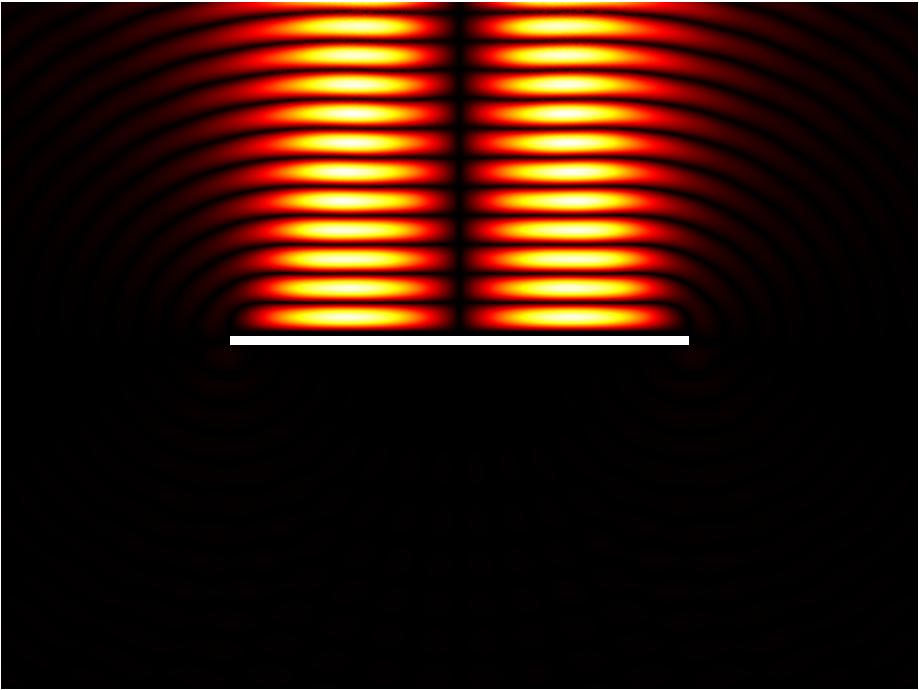}{0.3\textwidth}{(d) WS mode \#6}\label{fig:strip_tm_WS6} 
\fig{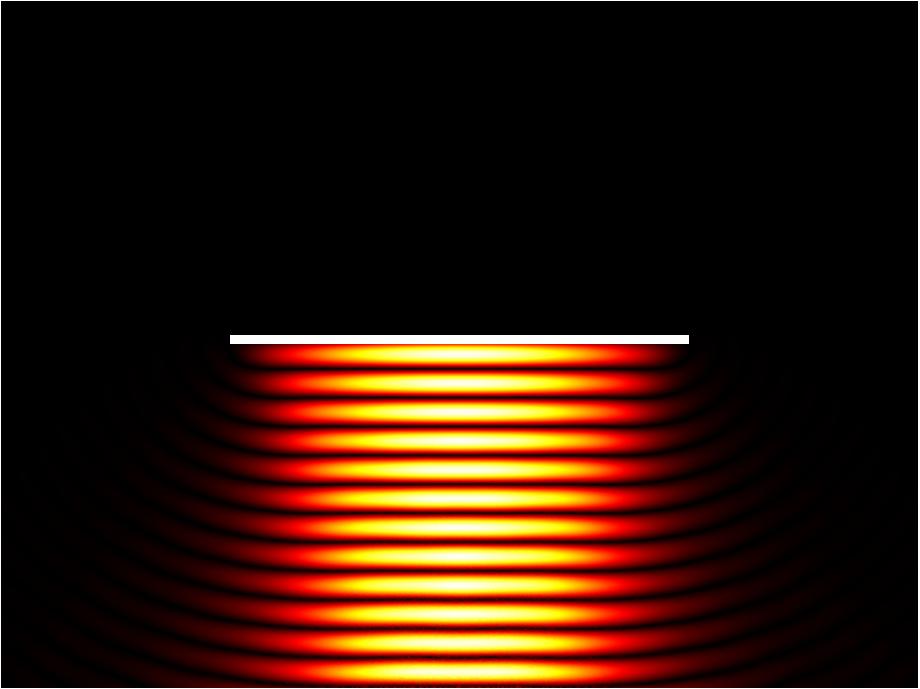}{0.3\textwidth}{(e) WS mode \#20}\label{fig:strip_tm_WS20}
\fig{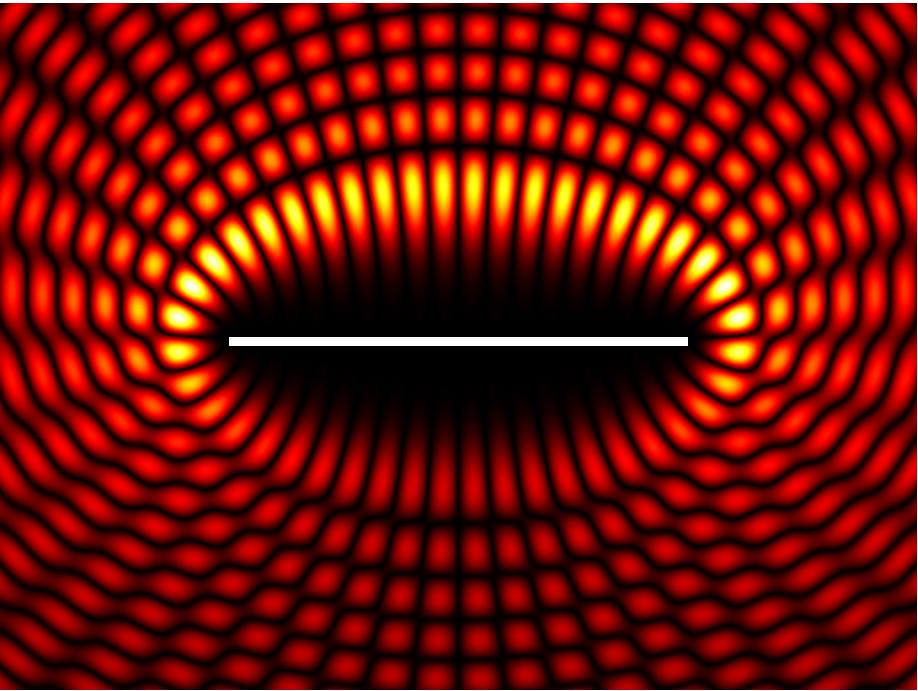}{0.3\textwidth}{(f) WS mode \#50}\label{fig:strip_tm_WS50}}
\caption{Selected WS modes of the sound-soft rectangular strip.}
\label{fig:WSmodes_soundsoft}
\end{figure*}

\begin{figure*}
\figline{\fig{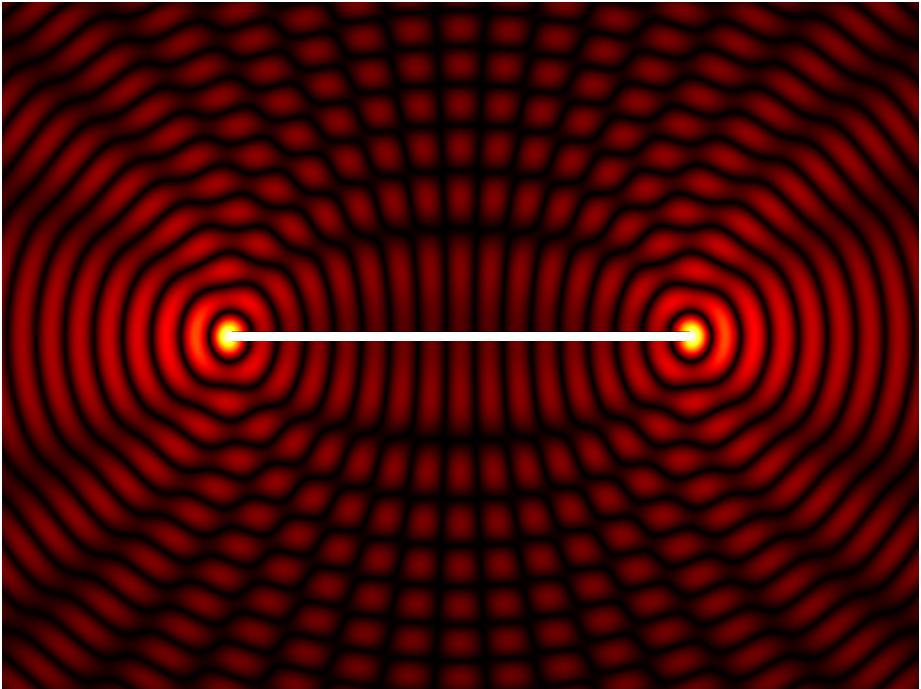}{0.3\textwidth}{(a) WS mode \#1}\label{fig:strip_te_1} 
\fig{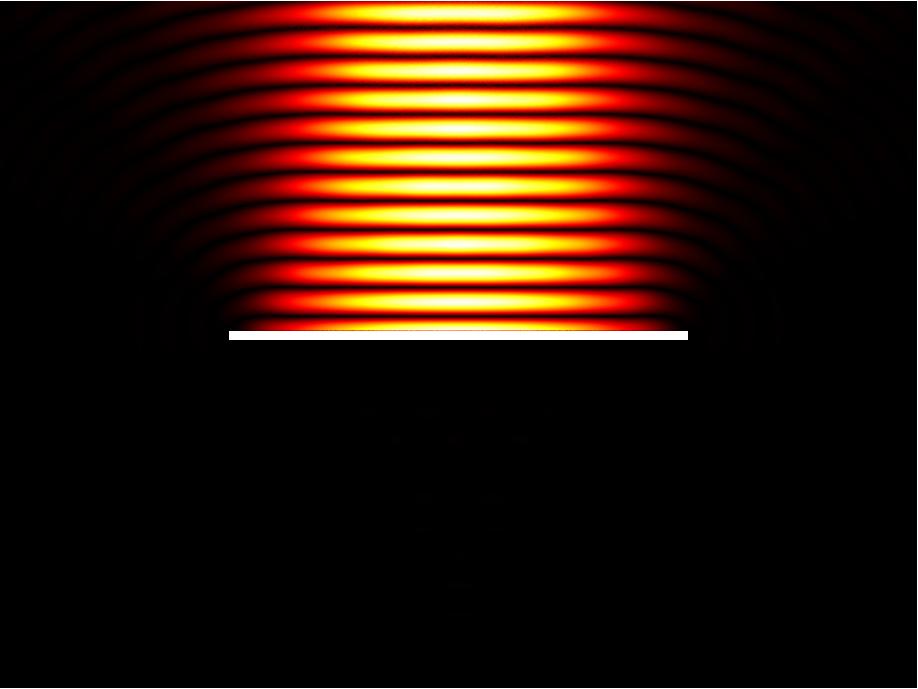}{0.3\textwidth}{(b) WS mode \#3}\label{fig:strip_te_3}
\fig{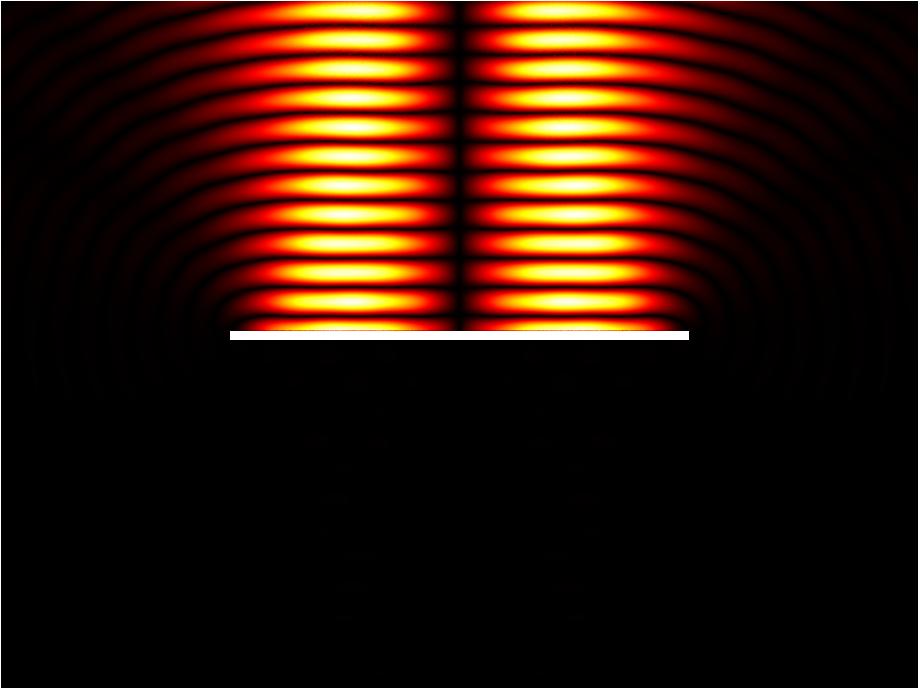}{0.3\textwidth}{(c) WS mode \#4}\label{fig:strip_te_4}}
\figline{\fig{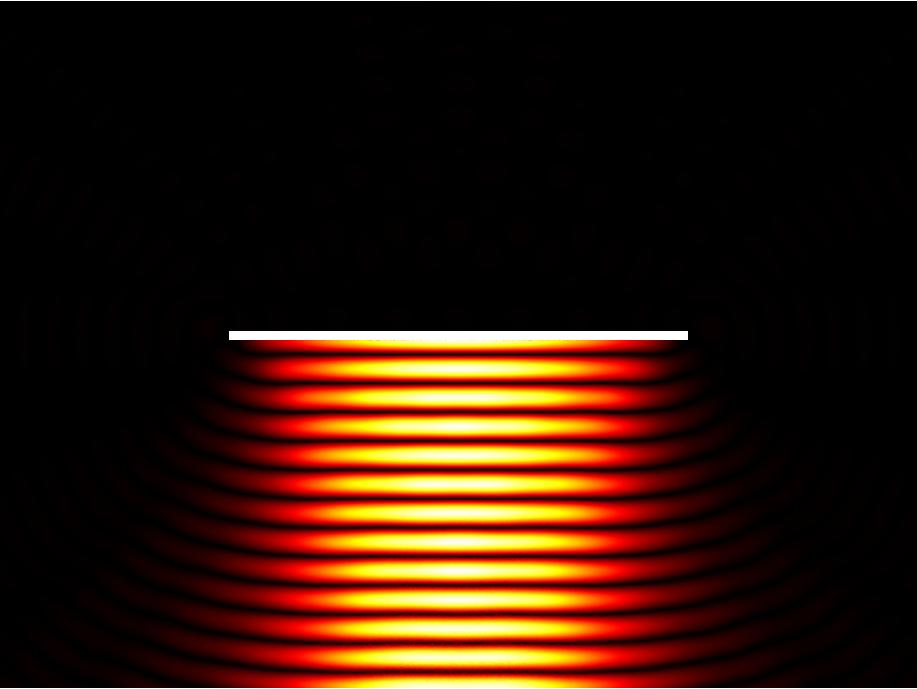}{0.3\textwidth}{(d) WS mode \#17}\label{fig:strip_te_107} 
\fig{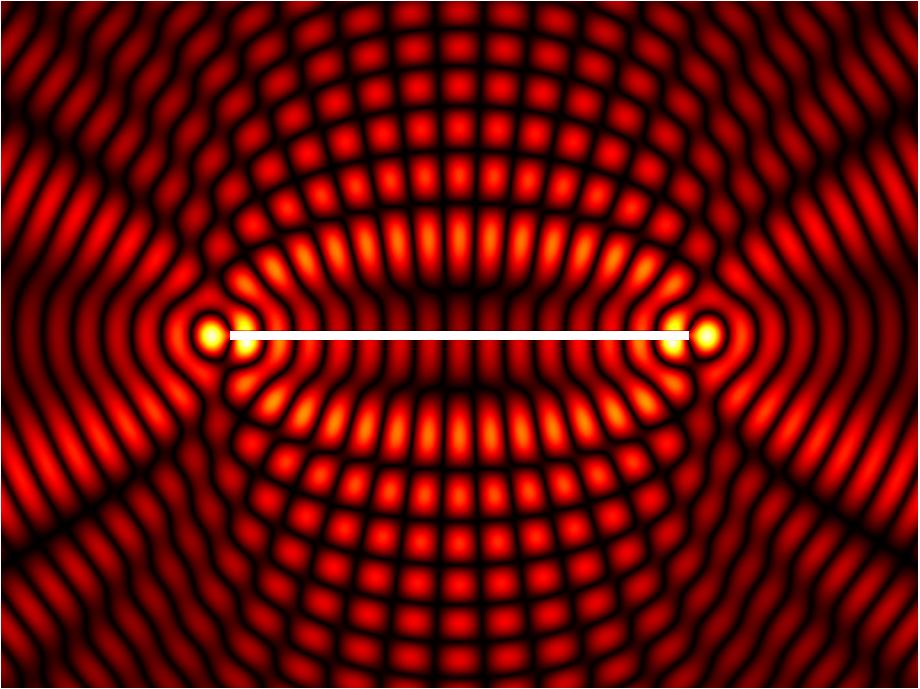}{0.3\textwidth}{(e) WS mode \#108}\label{fig:strip_te_108}
\fig{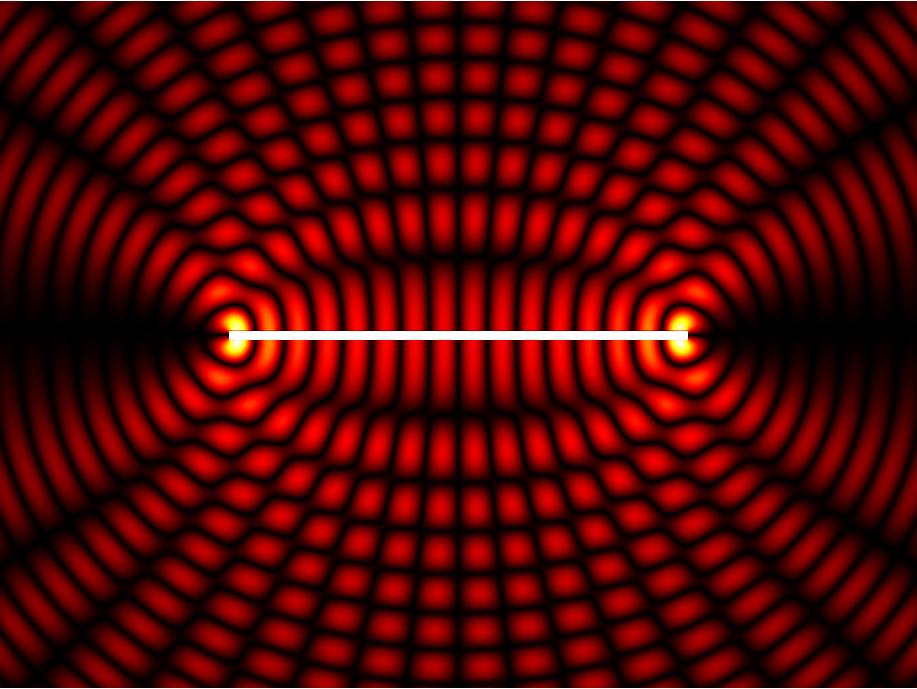}{0.3\textwidth}{(f) WS mode \#111}\label{fig:strip_te_111}}
\caption{Selected WS modes of the sound-hard rectangular strip.}
\label{fig:WSmodes_soundhard}
\end{figure*}

\begin{figure}[t]
\includegraphics{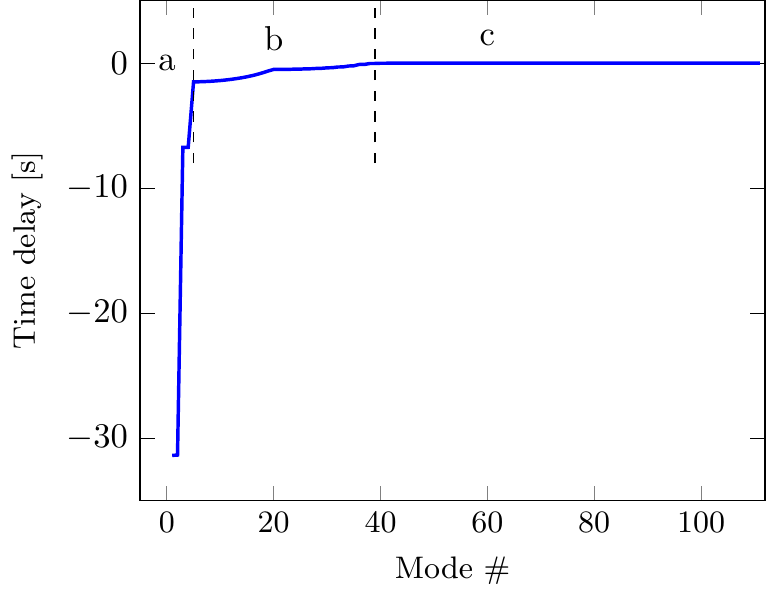}
\caption{WS time delays for the sound-soft rectangular strip.}
\label{fig:strip_tm_timedelays}
\end{figure}

\begin{figure}[t]
\includegraphics{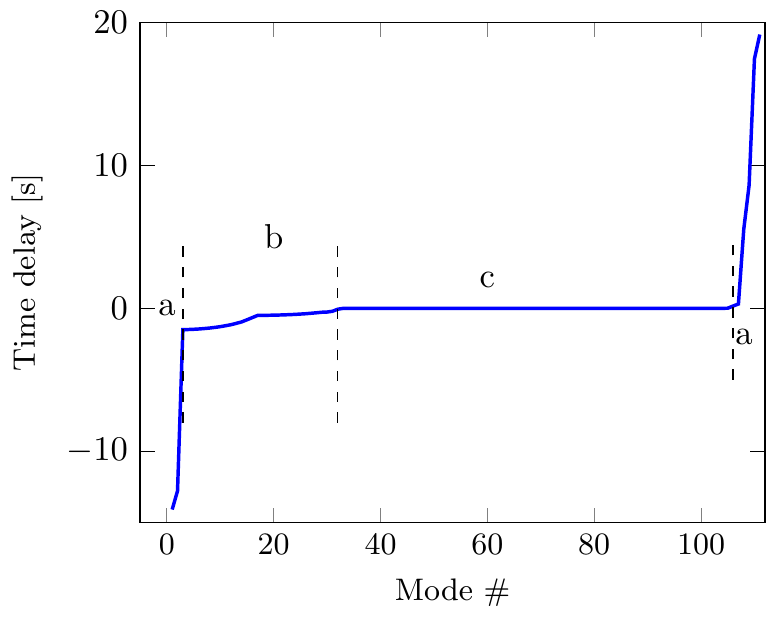}
\caption{WS time delays for the sound-hard rectangular strip.}
\label{fig:strip_te_timedelays}
\end{figure}

First, consider scattering from a 2D rectangular strip of size $50\mathrm{m} \times 1\mathrm{m}$ centered about $(0,0.25\mathrm{m})$. 
Note that the placement of the strip relative to the origin affects the WS modes as $\matr{Q}$ is related to $\matr{S}$ and $\matr{S}'$, both of which are origin-dependent. 
The strip is illuminated by $M=111$ incoming harmonics with $k=1$. 
Any other incoming field can be expressed as a superposition of these harmonics.
Knowledge of the scattered field associated with these illuminations is used to construct the strip's dense $111\times 111$ $\matr{Q}$ and $\matr{S}$ matrices, and $\matr{Q}$ is diagonalized as $\matr{Q} = \matr{W}\widebar{\matr{Q}} \matr{W}^\dag$. 
Figs.~\ref{fig:WSmodes_soundsoft} and \ref{fig:WSmodes_soundhard} show total fields associated with selected WS modes constructed by weighing incoming harmonics with entries of columns of $\matr{W}$ for sound-soft and sound-hard scatterers, respectively.  The corresponding time delays (diagonal elements of $\widebar{\matr{Q}}$) are shown in Figs.~\ref{fig:strip_tm_timedelays} and \ref{fig:strip_te_timedelays}. 
In what follows, WS modes are ordered by their time-delays (small to large) and indexed accordingly. \\

\noindent
\underline{Sound-Soft Scatterer}:
The modes can be classified into three groups:
\begin{enumerate}[a., leftmargin=* ]
    \item \emph{Corner Modes}: WS modes 1-4 illuminate the corners of the strip; total fields for modes 1 and 3 are shown in Figs.~\ref{fig:WSmodes_soundsoft}a and \ref{fig:WSmodes_soundsoft}b. 
    Modes 1 and 2 illuminate the two corners on the left (or right) side of the strip with a common-mode excitation (same magnitude and phase) while modes 3 and 4 excite the same with a differential mode excitations (same magnitude but opposite phase).
    Corner modes experience negative time delays (between $-32\mathrm{s}$ and $-6 \mathrm{s}$ in this case), because they spend less time in the system than they would in the absence of the scatterer.
    \item \emph{Ballistic Modes}:
    WS modes $5$-$39$ represent beam-like ballistic (or geometric optics) modes that specularly reflect off the strip. 
    Distributions of total fields for modes 5, 6, and 20 are shown in Figs. \ref{fig:WSmodes_soundsoft}c, \ref{fig:WSmodes_soundsoft}d, and \ref{fig:WSmodes_soundsoft}e, respectively. 
    Note that these modes avoid exciting the corners of the strip.
    Ballistic modes experience small negative time delays because these modes impinge on the strip from the $+\unitup{y}$ or $-\unitup{y}$ direction, reflecting off the strip's top or bottom surfaces and spending between $0s$ to $1.5\mathrm{s}$ less in the system depending on the angle of incidence compared to waves that do not interact with the scatterer (distance from origin to the top and bottom surface of the strip is $0.75\mathrm{m}$ and $0.25\mathrm{m}$, respectively). 
    \item \emph{Non-propagating Modes}: WS modes 40-111 do not excite the scatterer. The total field distribution for one such mode is shown in Fig.~\ref{fig:WSmodes_soundsoft}f. These modes experience near-zero time delays.
\end{enumerate}

\noindent
\underline{Sound-Hard Scatterer}: WS modes for a sound-hard strip likewise can be categorized into three groups:
\begin{enumerate}[a., leftmargin=*]
\item \emph{Corner/Surface Wave Modes}: Modes 1-2 and 106-111 excite both the corners and the edges of the strip. 
The distributions of the total fields when the strip is excited by WS modes 1, 108, and 111 are shown in Figs.~\ref{fig:WSmodes_soundhard}a, \ref{fig:WSmodes_soundhard}e, and \ref{fig:WSmodes_soundhard}f, respectively. 
Modes 1 and 2 only weakly excite the long edges of the strip as most of their energy rapidly scatters off the corners, resulting in negative time delays. 
Modes 106-111, in contrast, strongly excite the long edges of strip and have large positive time delays (and stored energy) as they involve surface waves that travel back and forth along the strip. Note that these surface wave phenomena and the associated positive time delays are unique to the sound-hard case and absent in the sound-soft case.
\item \emph{Ballistic Modes}: WS modes 3-32 represent beam-like ballistic modes that specularly reflect off the strip. Figs. \ref{fig:WSmodes_soundhard}b, \ref{fig:WSmodes_soundhard}c, and \ref{fig:WSmodes_soundhard}d show the distribution of the total field when the strip is illuminated by ballistic modes 3, 4, and 17, respectively. 
The behavior of these modes is nearly identical to that of ballistic modes for the sound-soft scatterer both in terms of the WS time delays and the distribution of total field. 
\item \emph{Non-propagating Modes}: WS modes 33-105 are non-propagating modes with near-zero time delays. These modes behave similarly to those for the sound-soft case. 
\end{enumerate}
This example demonstrates that WS modes can untangle fields into components associated with canonical scattering phenomena associated with well-defined time delays/dwell times and system energies.

\subsection{Open Cavity}

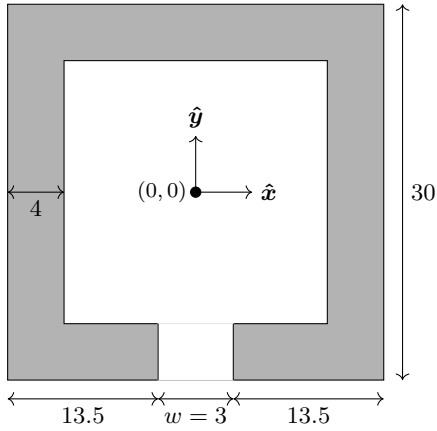
\begin{figure}
\centering
\begin{tikzpicture}

\filldraw [fill=black!30, draw=black] (0,0) rectangle (5,5);
\filldraw [fill=white, draw=black] (0.75,0.75) rectangle (4.25,4.25);
\filldraw [fill=white, draw=white] (2.0,0.0) rectangle (3,0.75);
\draw[draw=black] (2.0,0.0) -- (2,0.75);
\draw[draw=black] (3.0,0.0) -- (3.0,0.75);

\draw[<->] (0,-0.25) -- (2.0,-0.25);
\node at (1.0, -0.25) [below] {$13.5$};
\draw[<->] (2,-0.25) -- (3.0,-0.25);
\node at (2.5, -0.25) [below] {$w=3$};
\draw[<->] (3,-0.25) -- (5,-0.25);
\node at (4, -0.25) [below] {$13.5$};
\draw[<->] (5.25,0) -- (5.25,5);
\node at (5.25, 2.5) [right] {$30$};

\draw[<->] (0,2.5) -- (0.75,2.5);
\node at (0.375, 2.5) [below] {$4$};

\begin{scope}[shift={(2.5,2.5)}]
\draw[->]  (0,0) -- (0,0.75);
\draw[->]  (0,0) -- (0.75,0);
\node at (0.75,0) [right] {\small $\unitup{x}$};
\node at (0,0.75) [above] {\small $\unitup{y}$};
\node at (0,0) [left] {\footnotesize $(0,0)$};

\draw[fill=black] (0,0) circle (.5ex);
\end{scope}
\end{tikzpicture}
\caption{Dimensions (in $\mathrm{m}$) of an open cavity.}
\label{fig:cavity_geometry}
\end{figure}
\begin{figure*}[htb]
\figline{\fig{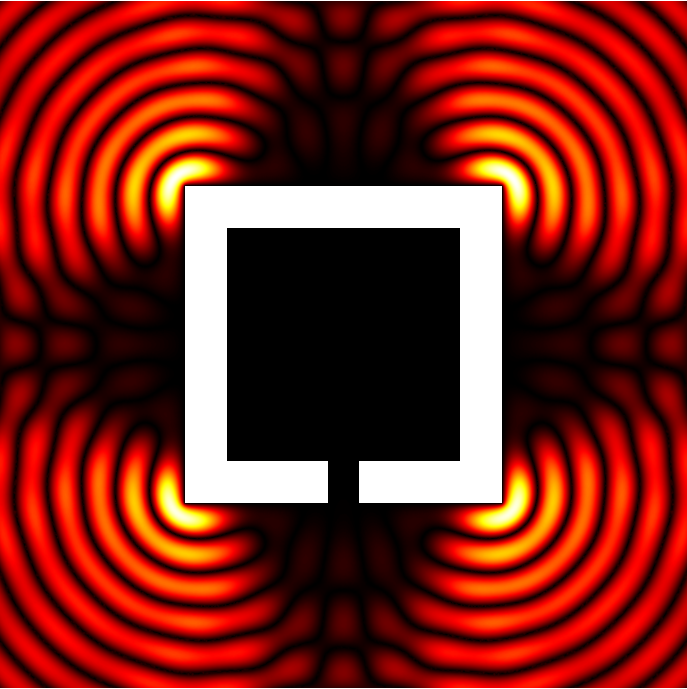}{0.3\textwidth}{(a) WS mode \#1}\label{fig:cavity_tm_WS1} 
\fig{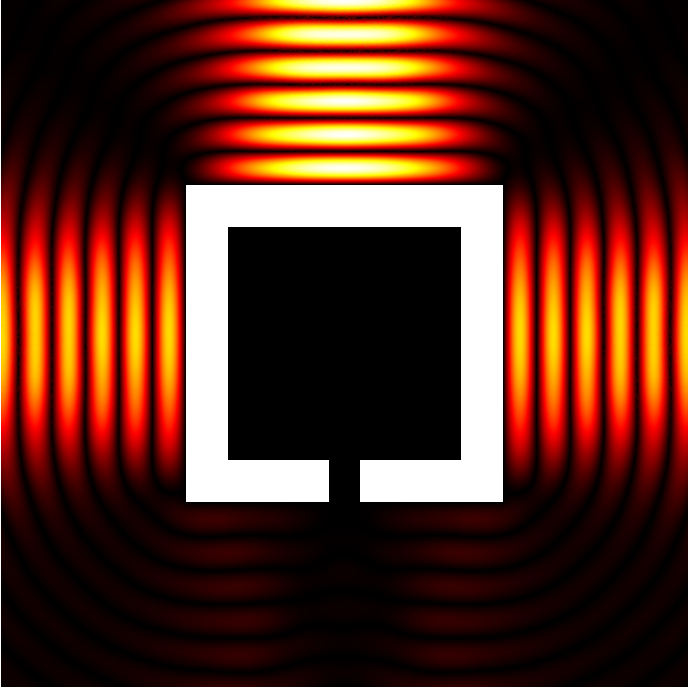}{0.3\textwidth}{(b) WS mode \#5}\label{fig:cavity_tm_WS5}
\fig{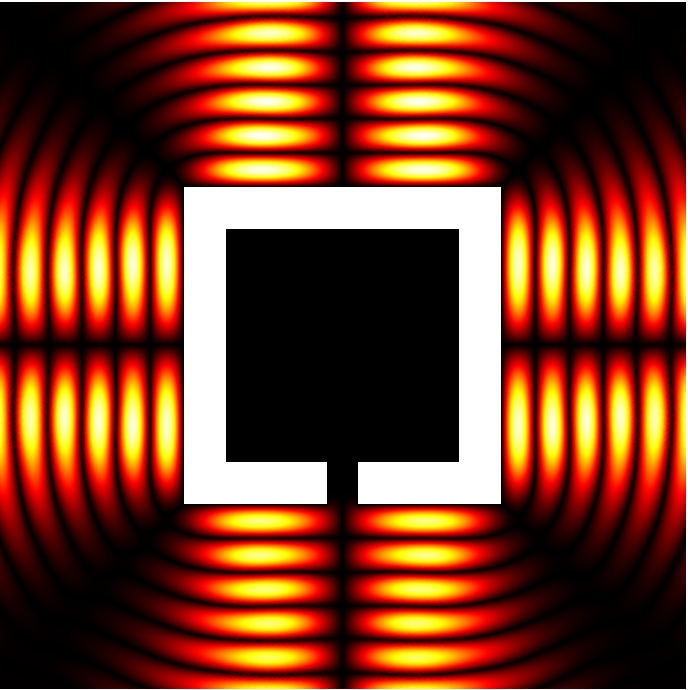}{0.3\textwidth}{(c) WS mode \#8}\label{fig:cavity_tm_WS8}}
\figline{\fig{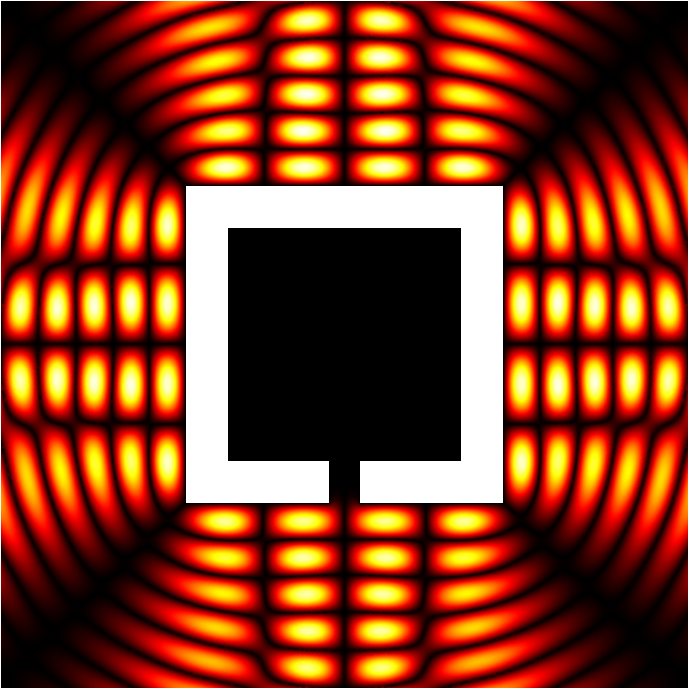}{0.3\textwidth}{(d) WS mode \#16}\label{fig:cavity_tm_WS16} 
\fig{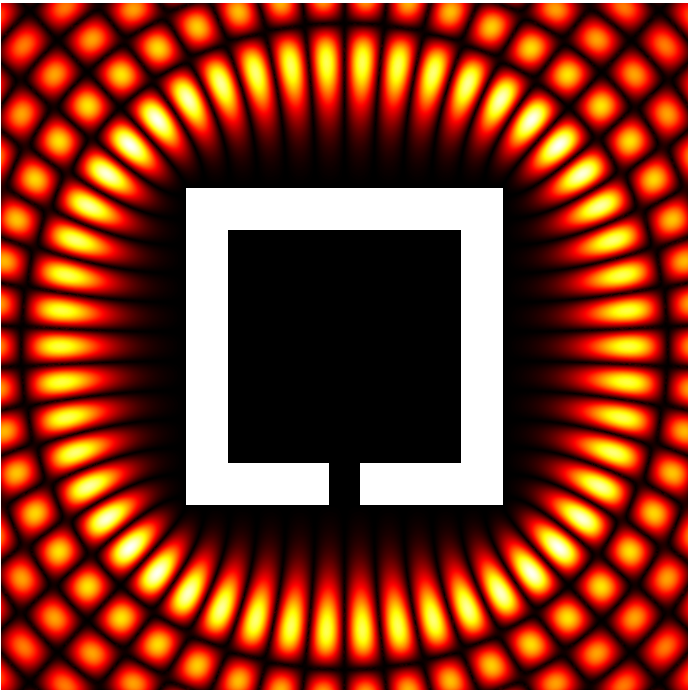}{0.3\textwidth}{(e) WS mode \#50}\label{fig:cavity_tm_WS50}
\fig{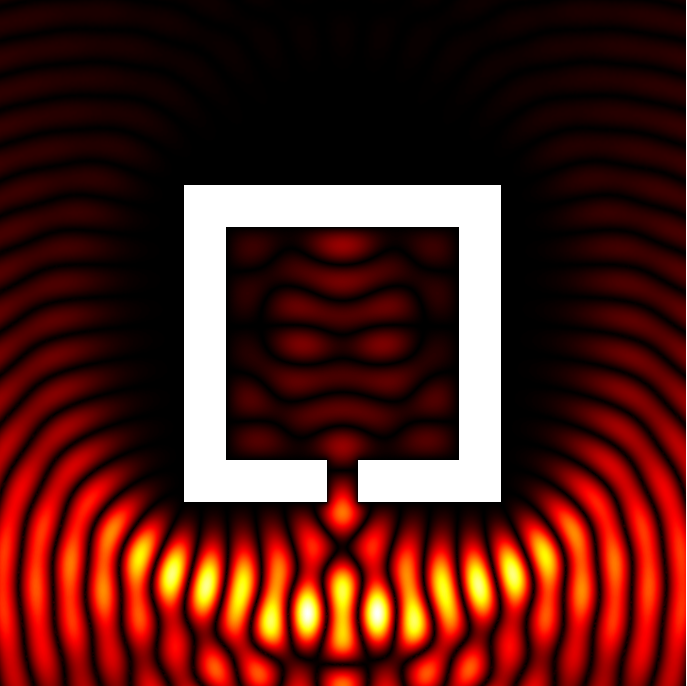}{0.3\textwidth}{(f) WS mode \#71}\label{fig:cavity_tm_WS71}}
\caption{Selected WS modes of a sound-soft cavity.}
\label{fig:WSmodes_cavity_TM}
\end{figure*}
\begin{figure*}[htb]
\figline{\fig{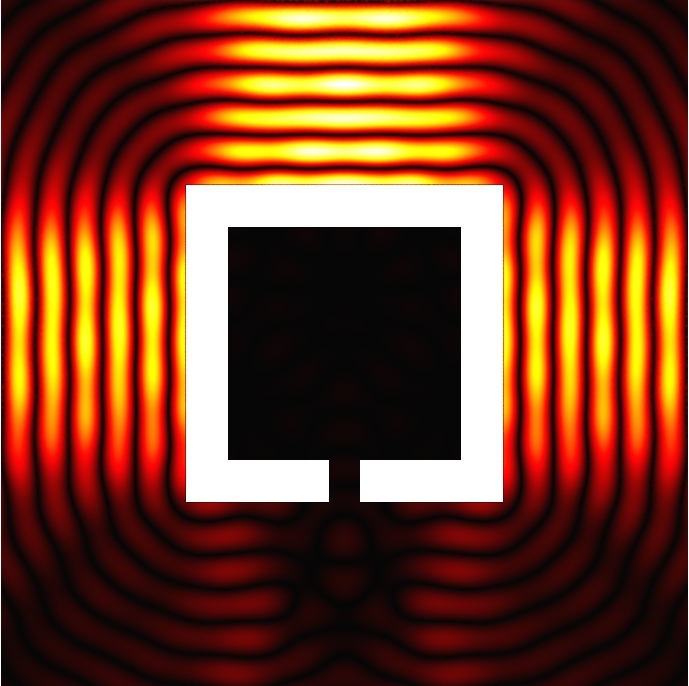}{0.3\textwidth}{(a) WS mode \#1}\label{fig:cavity_te_WS1} 
\fig{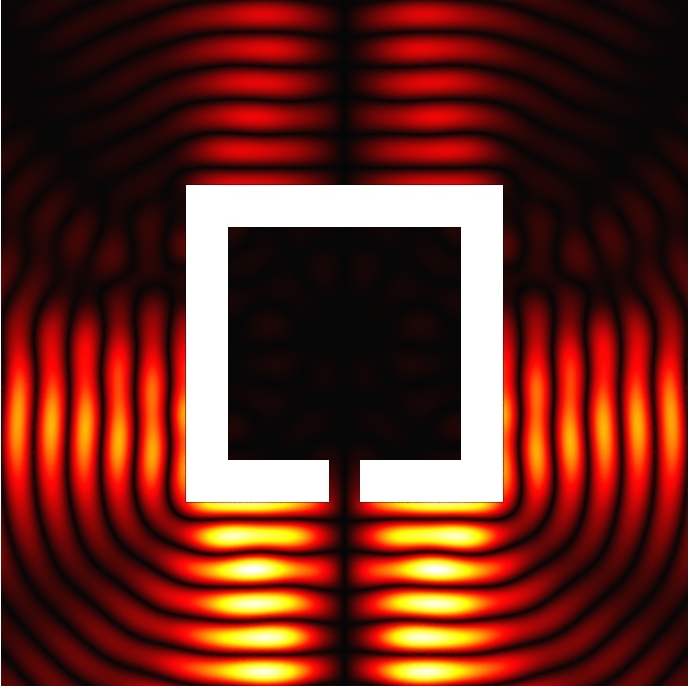}{0.3\textwidth}{(b) WS mode \#3}\label{fig:cavity_te_WS3}
\fig{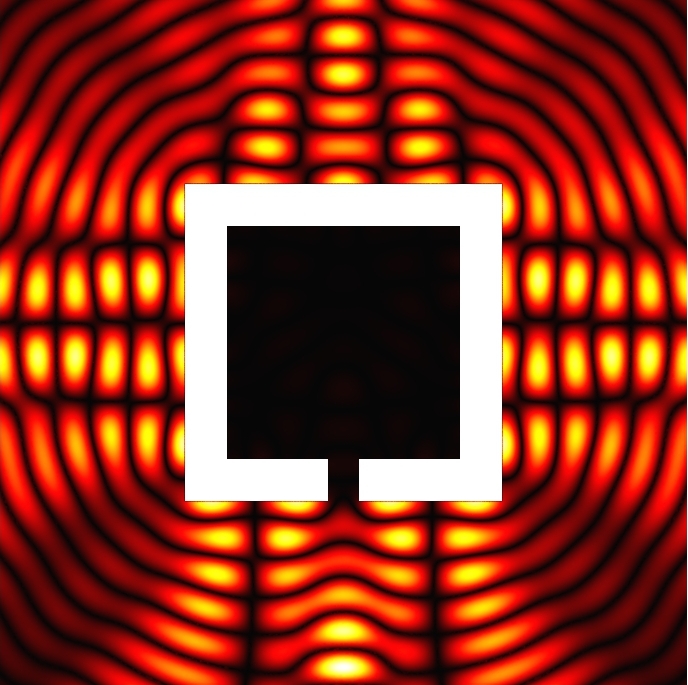}{0.3\textwidth}{(c) WS mode \#15}\label{fig:cavity_te_WS15}}
\figline{\fig{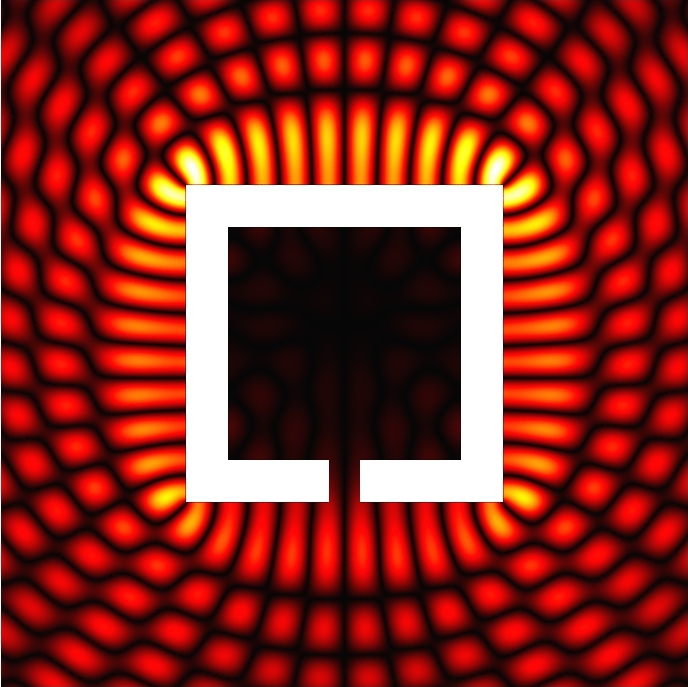}{0.3\textwidth}{(d) WS mode \#69}\label{fig:cavity_te_WS69} 
\fig{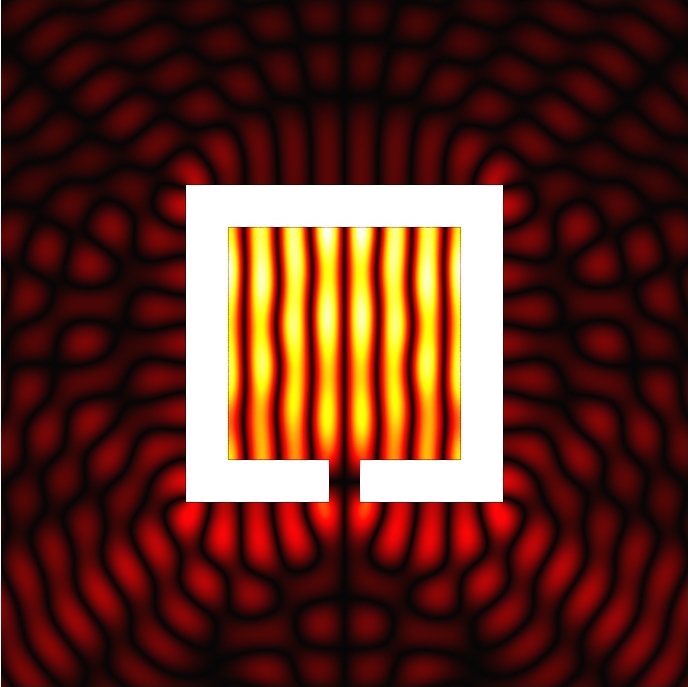}{0.3\textwidth}{(e) WS mode \#70}\label{fig:cavity_te_WS70}
\fig{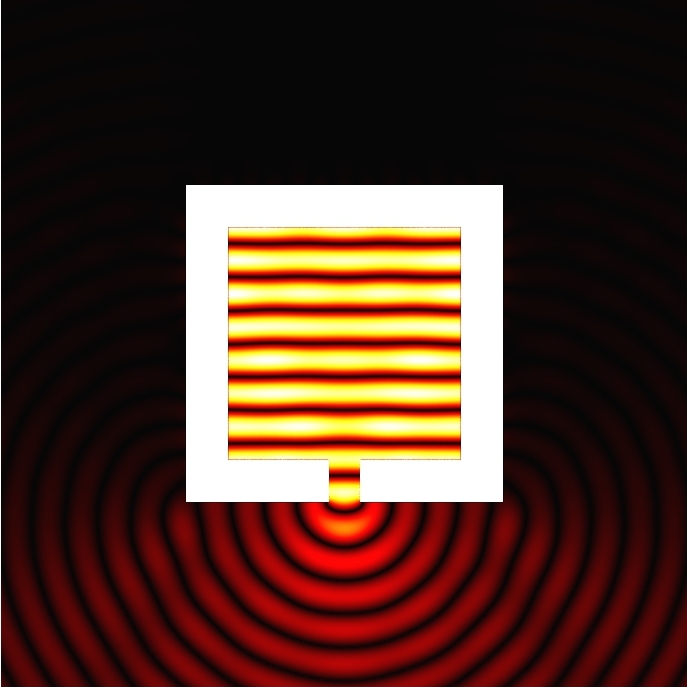}{0.3\textwidth}{(f) WS mode \#71}\label{fig:cavity_te_WS71}}
\caption{Selected WS modes of the sound-hard cavity.}
\label{fig:WSmodes_cavity_TE}
\end{figure*}
\begin{figure}[htb]
\includegraphics{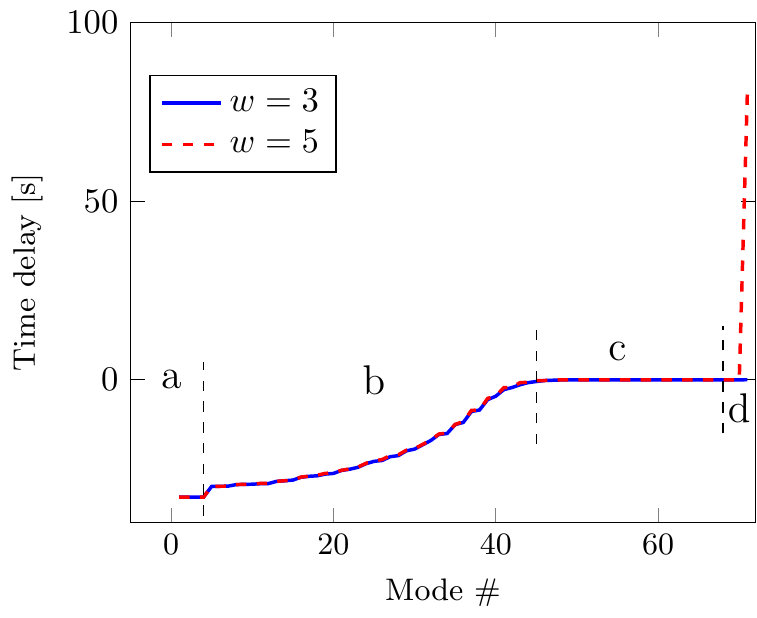}
\caption{WS time delays for the sound-soft cavity.}
\label{fig:cavity_tm_timedelays}
\end{figure}
\begin{figure}[htb]
\includegraphics{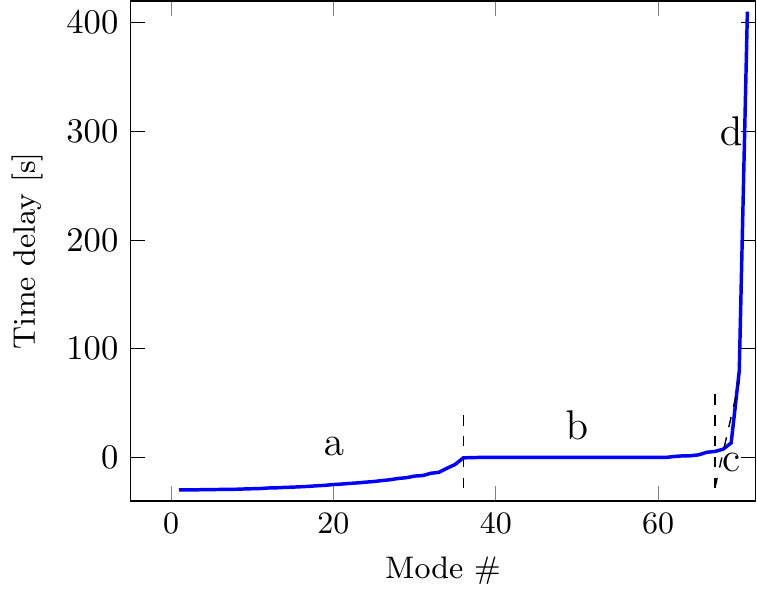}
\caption{WS time delays for the sound-hard cavity.}
\label{fig:cavity_te_timedelays}
\end{figure}

Consider scattering from the open cavity shown in Fig.~\ref{fig:cavity_geometry}. 
The cavity is illuminated with $M=71$ incoming harmonics with $k=1$. 
Knowledge of the scattered field is used to construct the $71 \times 71$ $\matr{S}$ and $\matr{Q}$ matrices. Next $\matr{Q}$ is diagonalized and the total fields associated with selected WS modes for cavities with sound-soft and sound-hard boundaries are shown in Figs.~\ref{fig:WSmodes_cavity_TM} and \ref{fig:WSmodes_cavity_TE}, respectively. Their corresponding time delays are shown in Figs.~\ref{fig:cavity_tm_timedelays} and \ref{fig:cavity_te_timedelays}.
\\

\begin{figure}[htb]
\includegraphics[width=0.3\textwidth]{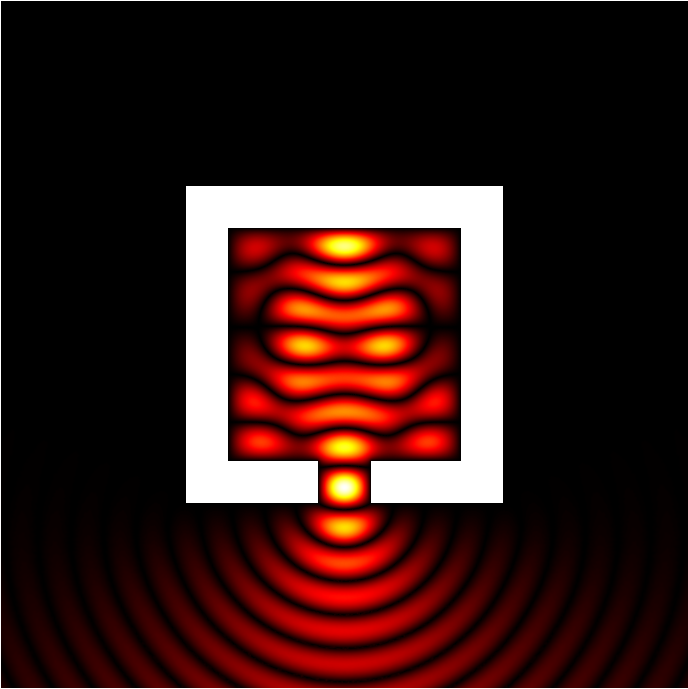}
\caption{WS modes 71 of the sound-soft cavity with the gap of $w=5~\mathrm{m}$.}
\label{fig:cavity_gap5}
\end{figure}


\noindent 
\underline{Sound-Soft Scatterer}: WS modes for the sound-soft cavity can be categorized into four groups:
\begin{enumerate}[a., leftmargin=*]
\item \emph{Corner Modes}: Modes 1-4 excite the corners of the cavity; the field distribution of WS mode 1 is shown in Fig.~\ref{fig:WSmodes_cavity_TM}a. Since these fields quickly leave the system after scattering off the corners, they have large negative time delays.
\item \emph{Ballistic Modes}: Modes 5-48 specularly reflect off the sides of the cavity without exciting the gap. Fields distributions for WS modes 5, 8, and 16 are shown in Figs.~\ref{fig:WSmodes_cavity_TM}b, \ref{fig:WSmodes_cavity_TM}c, and \ref{fig:WSmodes_cavity_TM}d, respectively. Note that, unlike for the rectangular strip, all edges are excited symmetrically. This is attributed to the cavity being placed symmetrically w.r.t. the origin.  
\item \emph{Non-propagating Modes}: Modes 49-70 do not excite the scatterer at all; an example field distribution is shown in Fig.~\ref{fig:WSmodes_cavity_TM}e.
These modes have near-zero time delays.
\item \emph{Cavity Modes}: Mode 71 excites the interior of the cavity. Fig.~\ref{fig:WSmodes_cavity_TM}f shows that when the gap width is $w=3~\mathrm{m}$, the cavity interior is not strongly excited. Indeed, the cavity's aperture can be considered as a short waveguide, which is under cutoff if $w=3~\mathrm{m}$. However, increasing the gap width to $w=5~\mathrm{m}$ results in strong fields inside the cavity, as shown in Fig.~\ref{fig:cavity_gap5}, because in this case the aperture supports one propagating mode. 
Cavity modes have large positive time delays and stored energies.
\end{enumerate}

\noindent
\underline{Sound-Hard Scatterer}: The WS modes can once again be categorized into four groups:
\begin{enumerate}[a., leftmargin=*]
\item \emph{Ballistic Modes}: WS modes 1-36 represent ballistic modes that specularly reflect off the outer walls of the cavity. Figs.~\ref{fig:WSmodes_cavity_TE}a, \ref{fig:WSmodes_cavity_TE}b, and \ref{fig:WSmodes_cavity_TE}c show the total fields when the sound-hard cavity is illuminated by modes 1, 3, and 15, respectively. 
These modes experience negative delays between $-29\mathrm{s}$ and $0\mathrm{s}$ depending on the angle of arrival of the incident field. 
\item \emph{Non-propagating Modes}: WS modes 37-68 are non-propagating modes that do not reach the scatterer. They have near-zero time delays.
\item \emph{Corner/Surface Wave Modes}: Modes 63-69 excite waves propagating along the exterior corners and side-walls of the cavity. 
The total field distribution for WS mode 69 is shown in Fig.~\ref{fig:WSmodes_cavity_TE}d. 
It can be seen that the fields are localized near, and launch from, the four corners and side-walls. These modes have small positive time delays. 
\item \emph{Cavity Modes}: Modes 70 and 71 are cavity modes. The total field distribution for these modes is shown in Figs.~\ref{fig:WSmodes_cavity_TE}e and \ref{fig:WSmodes_cavity_TE}f. For these WS modes, the incident field excites the cavity without exciting the corners or side-walls. These modes represent waves trapped in the cavity, have large stored energy, and experience large time delays.
Note that for the sound-hard boundary condition, the short aperture waveguide supports at least one mode even when $w=3~\mathrm{m}$, resulting in the strong excitation of the cavity interior for WS mode 71.
\end{enumerate}

\section{Conclusion}

This paper elucidated a WS theory for acoustic scattering problems governed by the Helmholtz equation involving sound-soft or sound-hard objects of arbitrary geometry.
The entries of the WS time delay matrix were cast in terms of renormalized volume integrals involving energy densities.  Numerical examples show that the eigenvectors of the of WS time delay matrix can be used to untangle scattering phenomena into canonical contributions characterized by well-defined energies and dwell times.

Current research extending the methodology in this paper is focused on
\begin{enumerate}
\item The development of fast frequency-sweep methods for computing a scatterer’s response to broadband excitations;
\item The design of acoustic systems and devices resulting in fields that exhibit desired group delays;
\item The phenomenological study of acoustic waves interacting with complicated structures and metamaterials.
\end{enumerate}

\appendix

\section{Modes for Scattering Systems}
\label{Appdix:modes}

Incoming fields satisfying the Helmholtz equation in 3D can be expressed as a superposition of incoming spherical waves
\begin{align}
    {\cal I}_{p}^i(\vrup) &= k j^{l+1} h_l^{(1)}(kr) {\cal X}_{lm}(\theta,\phi)
\end{align}
where $p=(l,m)$ for $l=0,...,L$ and $m=-l,...,l$, and $h_l^{(1)}(.)$ is the spherical Hankel function of type 1 and order $l$.
The spherical harmonic ${\cal X}_{p}(\theta,\phi)$ is defined as
\begin{align}
    {\cal X}_{p}(\theta,\phi) = (-1)^m \sqrt{\frac{2l+1}{4 \pi} \frac{(l-m)!}{(l+m)!}} P_l^m(\cos \theta) e^{j m \phi}
\label{eq:Ylm}
\end{align}
where $P_{l}^{m}(x)$ is the associated Legendre polynomial of degree $l$ and order $m$~\citep{Abr64}.
In addition to the orthogonality property in Eqn.~\eqref{eq:orthonormal}, the spherical harmonics satisfy
\begin{align}
    {\cal X}_{p}^*(\theta,\phi) &= (-1)^m {\cal X}_{\tilde{p}}
    \label{eq:conjugation}
\end{align}
where $\tilde{p} = (l,-m)$.
As $r\rightarrow \infty$, the incoming fields can be approximated as
\begin{align}
    \lim_{r\rightarrow \infty}{\cal I}^i_{lm}(r,\theta,\phi) \cong \frac{e^{jkr}}{r} {\cal X}_{lm}(\theta,\phi)
\end{align}
where the large argument approximation of spherical hankel function was used~\citep{}.

\section{Evaluation of the Surface Integrals}
\label{Appdix:sur_int}

Using the expressions for $\phi_{p,\infty}(\vrup)$, 
$\phi_{q,\infty}^*(\vrup)$,
$\phi_{p,\infty}'(\vrup)$, $\unitup{r} \cdot \nabla \phi_{q,\infty}^*(\vrup)$, $\unitup{r} \cdot \nabla \phi_{p,\infty}'(\vrup)$ on $d\Omega$, the surface integral on the LHS of \eqref{eq:5LHS} can be evaluated as $\frac{1}{2k}\left(I_1 - I_2 + I_3\right)$ where
\begin{subequations}
\begin{align}
    I_{1} &= \int_{d\Omega} \phi_{p,\infty}'(\vrup) \unitup{r} \cdot \nabla \phi_{q,\infty}^*(\vrup) d\vrup \\
    &= 2kR \delta_{pq}  - (-1)^m \left(j + kR \right)e^{2jkR} \matr{S}_{\tilde{p}q}^*   \nonumber \\
    &\quad - jk e^{-2jkR} \matr{S}_{qp}'   +  j k\sum_{m=1}^{M} \matr{S}_{mq}^* \matr{S}_{mp}'  \nonumber \\
    & \quad + (-1)^m \left( j -kR \right) e^{-2jkR}  \matr{S}_{q\tilde{p}}  \nonumber  \\
    I_{2} &=  \int_{d\Omega}    \phi_{q,\infty}^*(\vrup) \unitup{r} \cdot \nabla \phi_{p,\infty}'(\vrup) \\
    &= -2 k R \delta_{qp} - jk e^{-2jkR} \matr{S}'_{qp}   \nonumber \\
    &\quad - (-1)^m k R e^{-2jkR} \matr{S}_{q\tilde{p}}  - (-1)^m k R e^{2jkR} \matr{S}_{\tilde{p}q}^*  \nonumber \\
    & \quad - jk \sum_{m=1}^{M} \matr{S}_{mq}^* \matr{S}_{mp}'  \nonumber \\
    I_{3} &= \frac{1}{k} \int_{d\Omega} \unitup{r} \cdot \nabla \phi_{q,\infty}^*(\vrup) \phi_{p,\infty}(\vrup) d\vrup \\
    &= -(-1)^m j e^{-2jkR} \matr{S}_{q\tilde{p}}    + j(-1)^m e^{2jkR} \matr{S}_{\tilde{p}q}^*   \nonumber \,.
\end{align}
\end{subequations}
Here, indices are defined as $p=(l,m)$, $q=(l',m')$ and $\tilde{p} = (l,-m)$.
The evaluation of the above integrals was simplified by using the large argument approximation of Hankel functions and leveraging the unitarity of $\matr{S}$, i.e. $\sum_{m=1}^{M} \matr{S}_{mq}^* \matr{S}_{mp} = \delta_{q,p}$ .

\bibliographystyle{plain}
\bibliography{ref}






\end{document}